\documentclass[number,sort&compress]{elsarticle}
\usepackage[utf8]{inputenc}
\usepackage{hyperref}
\usepackage[nolist,nohyperlinks]{acronym}
\usepackage[british]{babel}
\usepackage{blindtext}
\usepackage{booktabs}
\usepackage{todonotes}
\usepackage{array}
\usepackage{booktabs}
\usepackage{tabularx}
\usepackage{subfig}
\usepackage{enumitem}

\usepackage{tikz}
\usetikzlibrary{arrows.meta}
\usetikzlibrary{positioning}
\usetikzlibrary{shapes.geometric}
\usetikzlibrary{trees}

\newcommand{\expnumber}[2]{{#1}\mathrm{e}{#2}}

\journal{Future Generation Computer Systems}

\date{}

\begin{acronym}
\acro{sna}[SNA]{Social Network Analysis}
\acro{isis}[ISIS]{Islamic State of Iraq and Syria}
\acro{svm}[SVM]{Support Vector Machine}
\acro{api}[API]{Application Programming Interface}
\end{acronym}

\begin{document}

\begin{frontmatter}

\title{Statistical Analysis of Risk Assessment Factors and Metrics to Evaluate Radicalisation in Twitter}

\author{Raúl Lara-Cabrera\fnref{corresponding}}
\ead{raul.lara@uam.es}
\author{Antonio Gonzalez-Pardo}
\ead{antonio.gonzalez@uam.es}
\author{David Camacho}
\ead{david.camacho@uam.es}
\address{Computer Science Department\\ Universidad Autónoma de Madrid, Spain}
\fntext[corresponding]{Corresponding author}

\begin{abstract}
Nowadays, Social Networks have become an essential communication tools producing a large amount of information about their users and their interactions, which can be analysed with Data Mining methods. In the last years, Social Networks are being used to radicalise people. In this paper, we study the performance of a set of indicators and their respective metrics, devoted to assess the risk of radicalisation of a precise individual on three different datasets. Keyword-based metrics, even though depending on the written language, performs well when measuring frustration, perception of discrimination as well as declaration of negative and positive ideas about Western society and Jihadism, respectively. However, metrics based on frequent habits such as writing ellipses are not well enough to characterise a user in risk of radicalisation. The paper presents a detailed description of both, the set of indicators used to asses the radicalisation in Social Networks and the set of datasets used to evaluate them. Finally, an experimental study over these datasets are carried out to evaluate the performance of the metrics considered.
\end{abstract}


\end{frontmatter}

\section{Introduction}

As a social structure that consists of actors and the interrelationships between them, Social Networks are inherently related to the human being, although their study did not become popular until the 1930s. Nowadays, Social Networks have escalated and increased their size due to the penetration of the Internet in the society, hence simplifying their analysis due to the high opening of data as well as their high volume.

\ac{sna} emerged as a set of methods and tools for the analysis of social structures and the investigation of the social aspects of these structures~\cite{scott2012social}. As an interdisciplinary research field, \ac{sna} together with other complex networks became a major paradigm in sociology as well as in other formal sciences~\cite{Borgatti892}.

An interesting application of \ac{sna} is the detection and extraction of behavioural patterns from data obtained from Social Networks. By combining metrics at different levels of the network (macroscopic, microscopic and mesoscopic) as well as by studying the evolution of relationships between individuals, behavioural models can be built that, somehow, represent the members of the network.

In the last years, the world in general and the West in particular are in a situation of constant danger due to the terrorist attacks perpetrated by the so-called \ac{isis}. It is not surprising that the fight against terrorism has become one of the priority objectives of any country.

Jihadist terrorism has a distinct feature compared to other kind of terrorism: the way they radicalise and recruit their militants. People usually begin to be radicalised by contacting radical individuals or groups, which provides feel of belonging and social recognition. Moreover, radicals tend to publish a lot of information on the Internet, specifically in Social Networks such as Twitter, Tumblr, Facebook, Instagram and Youtube~\cite{UNODC2012,Thompson2011}.

Therefore, it is interesting to apply techniques of \ac{sna} to the data published by jihadist radicalised people to extract behavioural patterns that, in turn, help to assess the risk of radicalisation.

The goal of this paper is to study on a set of indicators devoted to risk radicalisation assessment and their respective metrics used to quantify them. These indicators come from several experts psychologists on radicalisation and are focused on the behaviour expressed in Social Networks. Furthermore, these indicators are easy to compute in an automated way, gathering and analysing messages from Social Networks. This should improve the present radicalisation risk assessment process, which is frequently done using manual tools such as VERA~\cite{pressman2012calibrating,Pressman2016}.

The main contributions of the present paper are the following:
\begin{itemize}
\item Five indicators to assess the risk of radicalisation as well as the metrics to quantify them are presented and studied.
\item Three datasets have been built in order to analyse the performance of the aforementioned metrics.
\item The results obtained provide guidance on how to improve the present indicators in order to achieve better performance when assessing risk of radicalisation in Social Networks.
\end{itemize}

The remainder of the paper is structured as follows. First a brief introduction of the principles of complex networks and what are the most commonly used algorithm to analyse these networks can be found in Section \ref{sec:background}. Then, the different indicators and the datasets used in this work are described in Section \ref{sec:materials}. Section \ref{sec:analysis} provides a detail description of the experimental phase that evaluate the performance of the metrics.Finally, the conclusions drawn from the experimental phase and future research works can be found in Section \ref{sec:conclusions}.

\section{Background}
\label{sec:background}

Complex networks have been studied mainly in the form of mathematical graph theory and also in the social sciences~\cite{doi:10.1137/S003614450342480}. Nowadays, the focus have shifted to networks with a high number of vertices and edges, trying to analyse their large-scale statistical properties. The presentation of models for scale-free~\cite{Watss1998} and power-law~\cite{Barabasi15101999}, as well as the study of new algorithms related to complex networks such as clustering~\cite{Bello2011,Schaeffer200727} and community finding~\cite{Bello2012,Malliaros201395,2017-GonzalezFGCS}, have led to the study of many different issues in this sense.

There are many works related to the study of different types of complex networks in the literature. For instance, the information network made of citations between academic papers in which every node is a paper and there is a directed edge if a paper cites another~\cite{egghe1990introduction}. Other approach is the one followed by ~\cite{Lara-Cabrera2014b,CottaG07,Barabasi2002590}, in these works the authors of the academic papers are represented in the nodes of the co-authorship network and the different edges imply there is at least an academic paper written by them. Regarding Social Networks~\cite{Scott1988}, there are studies on business relationships between companies~\cite{mizruchi1982american}, supply chain context~\cite{JSCM:JSCM03166} and obesity~\cite{Valente2009202}, to name a few.

Concerning terrorism and Social Networks, there are several approaches in the literature to tackle this problem. For instance, \citet{Wadhwa2013} proposed a Data Mining approach for detecting the dynamic behaviour of radicals by analysing the messages posted. The approach consists of a message filtering and preprocessing stage followed by a topic identification algorithm, and other \ac{sna} techniques such as community detection and identification of key nodes.

\citet{OCallaghan2014} performed a detailed analysis of the communities found in messages and videos from Twitter and Youtube regarding the Syria conflict. Their findings indicate that social media activity in Syria was more complicated than that reported by other studies. Furthermore, the authors studied the effect that certain key events had over the number of videos uploaded to Youtube.

Expanding the focus, \citet{Ferrara2016} designed a Machine Learning approach to detect extremist users levering temporal, network and meta-data features. In addition, this approach is able to predict content adopters and interaction reciprocity in social media. A similar approach was used by \citet{Agarwal2015a}, who used a \ac{svm} classifier to detect online radicalisation on Twitter, following a semi-supervised learning scheme. Furthermore, \citet{Ashcroft2016} followed a Machine Learning approach as well, in order to classify a tweet as containing material that is supporting jihadists groups or not. For its part, \citet{Kaati2016} used the AdaBoost classifier to detect tweets that disseminates jihadist propaganda through the social media.

On the other hand, \citet{Glowacki2016} show that the formation of raids for intergroup violence depends on the presence of specific leaders who tend to occupy a central position in the social network driven by the friendship relationship.

For a more comprehensive review of the literature, \citet{Correa2013} published a survey on solutions to detect and analyse online radicalisation.

\section{Materials and methods}\label{sec:materials}
This section is devoted to describe a set of indicators to assess the risk of radicalisation as well as to explain the datasets we have used to study the performance of the aforementioned indicators.

\subsection{Indicators}
As previously mentioned, the goal of this paper is to study the efficiency of different indicators that highlight those Social Network users with high risk of being radicalised. The studied indicators come from several expert psychologists on radicalisation and terrorism that are used to assess the radicalisation risk of an individual. Although it is possible to use many indicators on this context (see~\cite{gilperez2017initial,pressman2012calibrating,Pressman2016}), we focused on a manageable set of indicators which, in turn, are easily measurable using Social Networks data.

The set consists of \textbf{five indicators} grouped in two categories: \textit{personality} and \textit{interpersonal} relationships, and \textit{attitudes} and \textit{beliefs} towards Muslim religion and Western society. The former contains those indicators related to the writing style specific for each user, whereas the later are indicators are measured by the content of the tweets:

\begin{itemize}
\item \textbf{Personality related Indicators}:
\begin{enumerate}[label=\textbf{I\arabic*}]
\item \textit{The individual is frustrated}. To measure this indicator in Social Networks we will take into account some aspects such as swearing, writing sentences fully capitalised and using words with negative content.
\item \textit{The individual is introverted}. Aspects to be analysed: using ellipses (i.e. \ldots) in the messages and measuring their length (introverted people tend to write short sentences).
\end{enumerate}
\item \textbf{Attitudes and beliefs related Indicators}:
\begin{enumerate}[label=\textbf{I\arabic*}]
\setcounter{enumi}{2}
\item \textit{Perception of discrimination for being Muslim}. This perception may be expressed in the messages using some keywords related to discrimination.
\item \textit{Expressing negative ideas about Western society}. As occurs with discrimination, it will be used several keywords related to negative ideas when anybody talks about the Western life style.
\item \textit{Expressing positive ideas about jihadism}. Radical people show support and positive ideas about those engaged in the Jihad. Again, it is possible to analyse the usage of keywords related to this issue.
\end{enumerate}
\end{itemize}

Please refer to~\cite{CabreraEtAl} for a deeper explanation on the indicators and their operativeness (how they can be measured in a quantitative way), and Table~\ref{tbl:keywords} for the initial set of keywords used for each indicator.

\begin{table}[ht]
    \renewcommand{\arraystretch}{1.3}
	\centering
	\begin{tabularx}{.9\textwidth}{@{}p{4cm}X@{}}
		\textbf{Indicator} & \textbf{Initial keywords}\\
        \toprule
        \textbf{I1}. The individual is frustrated. & shit, crap, damn, fuck.\\
       \textbf{I1} Use of words with negative content. & hate, guilt, shame, terrible, horrible, bad, fault.\\
        \textbf{I3}. Perception of discrimination for being Muslim. & Muslim, sick, hate, discrimination, people, racism, religion.\\
        \textbf{I4}. Expressing negative ideas about Western society. & western, hate, suck, people, west, europe, usa, US, bloody, sick, impure, kuffar, kafir.\\
        \textbf{I5}. Expressing positive ideas about jihadism. & islamic, state, caliphate, rise, mujahideen, mujahid, help, fight, weapon, gun, weapons.\\ \bottomrule
	\end{tabularx}
    \vspace{1em}
    \caption{Initial keywords used to evaluate each indicator, they have been expanded to a larger set of keywords with synonyms obtained from Wordnet, and by seeking the occurrences of the stem of the words (stemming process). This is a two-step process: first keywords are expanded with Wordnet and then we keep their stems, which is what the method look for in the tweets.}\label{tbl:keywords}
\end{table}

\subsection{Datasets}
\paragraph{D1}
Due to the special sensitiveness of this paper's topic, it is really difficult to find a publicly accessible dataset\footnote{All of the datasets used in this work will be publicly available. They can be downloaded from: http://aida.ii.uam.es/resources}. Authors in the literature usually build their datasets on their own and they rarely publish them. In a previous work~\cite{CabreraEtAl} we used a dataset that, as far as we know, is the only open published dataset including Twitter usernames and tweets from \ac{isis} sympathizers. Furthermore, this dataset was built and curated by a digital agency that serves government agencies and is available at Kaggle\footnote{\url{https://www.kaggle.com/kzaman/how-isis-uses-twitter} (Last accessed: April, 2017)} with the following description: ``We scraped over 17,000 tweets from 100+ pro-ISIS fan-boys from all over the world since the November 2015 Paris Attacks''. Hence, this dataset should be considered as the most reliable.

\paragraph{D2}
Other dataset was the one built by volunteers all around the world from the so-called Anonymous collective that gathered and published many Twitter accounts of people related to \ac{isis} during the \texttt{\#OpISIS} operation~\cite{nytimes}. Several dump files can be found at Pastebin\footnote{\url{https://pastebin.com/u/CyberRog} (Last accessed: April, 2017)}, a web platform frequently used to publish this kind of information. Unlike with the previous one, this dataset has not been validated by experts but it was the same anonymous users who reported the accounts in Twitter with a certain hashtag and somebody grouped them in a dump file. So the reliability of this dataset should be taken with caution. One of the  relevant characteristics of this dataset is realted to the number of languages used in the tweets, several languages as English, Arabic, Russian, Turkish, French, have been employed to write them. Most of the tweets have been written in Arabic language, whereas English language are used only in a minor proportion ($1.65\%$ between both US and UK English). Next Table \ref{tbl:languages}, shows a distribution of these tweets by language.

\begin{table}[ht]
\renewcommand{\arraystretch}{1.3}
\centering
\caption{Top 10 writing languages detected in Anonymous dataset.}
\label{tbl:languages}
\begin{tabular}{rlc}
  \hline
 & Language & Percentage ($\%$) \\ 
  \hline
  1 & Arabic & \textbf{88.98} \\ 
  2 & Russian & 4.45 \\ 
  3 & Turkish & 2.23 \\ 
  4 & US English & \textit{1.55} \\ 
  5 & French & 1.30 \\ 
  6 & Dutch & 0.91 \\ 
  7 & UK English & \textit{0.10} \\ 
  8 & Urdu & 0.08 \\ 
  9 & Bulgarian & 0.06 \\ 
  10 & Hindi & 0.05 \\ 
  \hline
\end{tabular}
\end{table}

Once the dump files were downloaded, a tool developed in Python downloaded information about users and their tweets by making requests to the Twitter API.\@ The design of this tool took into account the limitations of the \ac{api}, such as the maximum number of requests that is limited in time as well as the number of tweets that can be downloaded from a user in each request.

\paragraph{D3}
To complement the above and analyse the performance of the metrics computed from the indicators we built another dataset. This new dataset represents a set of tweets from users that were randomly selected (so it can be used to evaluate any algorithm or metric, comparing random data against the data target). More precisely, we attached a software crawler to the streaming \ac{api} of Twitter during 30 seconds and then we randomly selected 120 users. To avoid any bias on the dataset, we did not use any keyword nor hashtag in the query, because we were trying to obtain a dataset as randomised as possible. Finally, we gathered tweets published by the aforementioned users.

Table \ref{tbl:datasets} shows a summary of previous datasets, the number of users and total amount of tweets are shown. the number of users are quite equivalent for all the datasets considered (around a hundred), the number of tweets have been increased for the random dataset (D3). Finally, to better understood the structure of these networks, the average number of tweets and the standard deviation for each user have been included. As it could be expected, the random dataset has the higher average of tweets per user. This is a side effect of the limitations imposed by the Twitter's API: it is only possible to obtain, at most, up to 3200 of a user’s most recent Tweets, so it makes sense that the average number of tweets for a random sample is around the half point within that range.

\begin{table}[ht]
	\renewcommand{\arraystretch}{1.3}
	\centering
	\begin{tabularx}{.7\textwidth}{@{}lccc@{}}
		& D1 & D2 & D3 \\ \toprule
		Number of users & $112$ & $142$ & $120$ \\
		Number of tweets & $17410$ & $76286$ & $173530$ \\
        Avg. tweets per user & $155.45$ & $527.23$ & $1446.08$ \\
        Stdev. tweets per user & $269.54$ & $901.48$ & $1114.92$ \\ \bottomrule
	\end{tabularx}
	\caption{Summary of the datasets used in this work.}\label{tbl:datasets}
\end{table}

\section{Analysis of the metrics over the datasets} \label{sec:analysis}
This section studies and analyses the performance of the different indicators over the datasets. These indicators have been operationalised (measured) using several metrics, in form of specific kind of words as swear/negative ideas or keywords, which can be extracted from the tweets to be later statistically processed.

\subsection{Experimental Setup}
\begin{figure}[ht]
\centering
\resizebox{0.7\textwidth}{!}{
\begin{tikzpicture}[%
	>=stealth,
    auto,scale=0.8]
    \tikzstyle{block} = [rectangle, draw, text width=6em, 
                     	text centered, 
                     	minimum height=2em, node distance=0.5cm]
    \tikzstyle{blank} = [rectangle, node distance=1cm]
    \node[blank] (init) {};
    \node[block, below=of init] (cleaning) {\small Remove URLs and usernames};
    \draw[->](init) -- node[right]{\small messages} (cleaning);
    \node[block, below=of cleaning] (sent_tokenizer) {\small Sentence tokenizer};
    \draw[->](cleaning) -- (sent_tokenizer);
    \node[block, below=1.5 of sent_tokenizer] (stem) {\small Stemming};
    \draw[->](sent_tokenizer) -- node[right]{\small sentences} (stem);
    \node[block,left=of stem,fill=blue!10] (sentlength) {\small \textbf{I1} Sentence length, ellipses};
    \node[block,right=of stem,fill=blue!10](frustration) {\small \textbf{I2} Capitalization, Swear/Negative Words};
    \draw[->](sent_tokenizer.south) ++(0,-5mm) -| (sentlength.north);
    \draw[->](sent_tokenizer.south) ++(0,-5mm) -| (frustration.north);
    \node[block,below=2.5 of stem,fill=blue!10] (western) {\small \textbf{I4} Neg. Ideas Western Society};
    \node[block,left=of western,fill=blue!10] (jihadism) {\small \textbf{I5} Pos. Ideas Jihadism};
    \node[block,right=of western,fill=blue!10] (discrimination) {\small \textbf{I3} Discrimination};
    \draw[->](stem) -- node[right,yshift=8mm]{\small words} (western);
    \draw[->](stem.south) ++(0,-18mm) -| (jihadism.north);
    \draw[->](stem.south) ++(0,-18mm) -| (discrimination.north);
    \draw[->](stem.south) ++(0,-18mm) -| (frustration.south);
\end{tikzpicture}
}
\caption{Data flow: preprocessing, cleaning and analysis to compute indicators.}
\label{fig:process}
\end{figure}
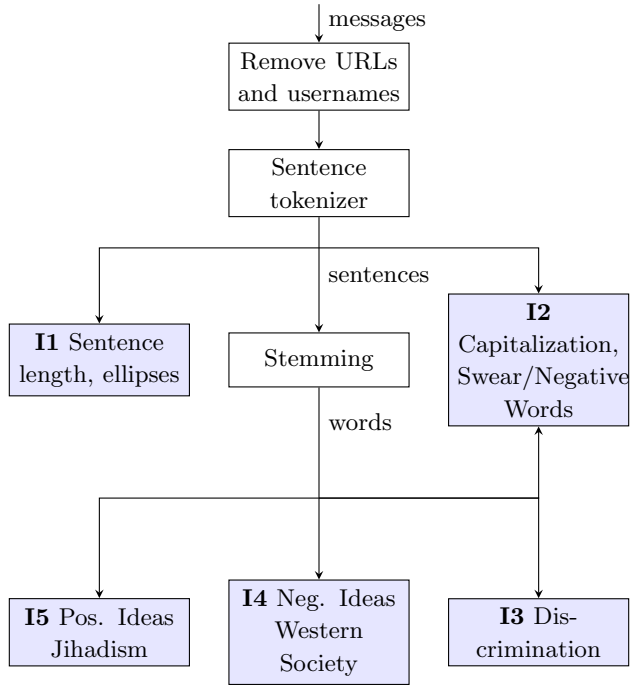

To study the performance of the indicators, we followed a process that involved several stages (see Figure~\ref{fig:process}). The first one, data preprocessing, homogenizes all the datasets as well as the messages from every user, removing the URLs and mentions using regular expressions. This is performed due to the low information this items provide to the indicators as they are defined.

As indicators \textbf{I1} and \textbf{I2} are based on sentences, messages are then tokenized into this unit of language by taking into account the punctuation marks and new lines found in the messages.

The remaining indicators (\textbf{I3}, \textbf{I4} and \textbf{I5}) are computed by counting keywords related to the indicator, we decided to widen the search by expanding the set of keywords with their respective synonyms querying WordNet~\cite{miller1995wordnet}, and by seeking the occurrences of the stem of the words (stemming stage). The latter process avoids unwanted situations as not counting as an occurrence the word and its plural form, and is generally known as stemming.

Once the metrics associated to the indicators were computed for the three datasets, it have been analysed the distribution of these values and also compared them among the datasets. Note that metrics are quantitative measures related to some indicator in a many-to-one relationship, thus an indicator may comprise several metrics. Following, the results for each indicator are described and analysed individually.

\subsection{Frustration}
To measure the frustration of any user we focused on two metrics: \textit{swearing} and the usage of \textit{words with negative content}. They are computed by counting the frequency of their respective keywords, that is, the number of times a keyword appears in a tweet.

\begin{figure}[ht]
	\centering
	\subfloat[][]{\includegraphics[width=.5\textwidth]{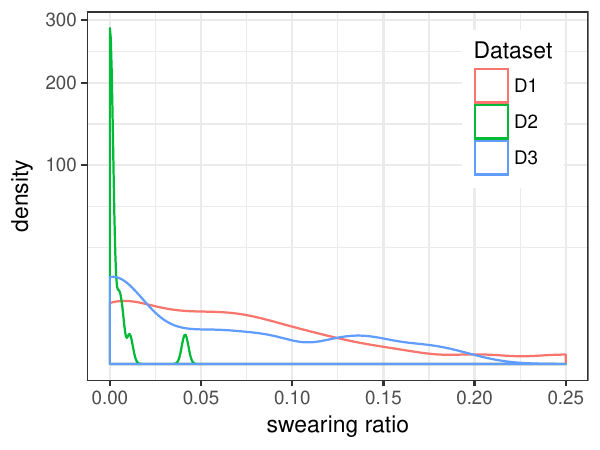}\label{fig:ecdf_swearing}}
	\subfloat[][]{\includegraphics[width=.5\textwidth]{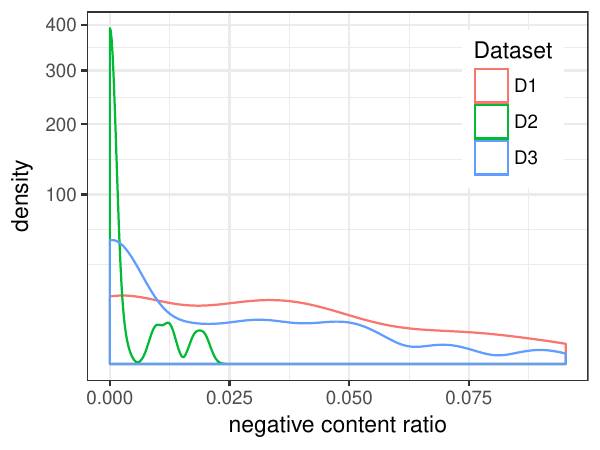}\label{fig:ecdf_negative}}
	\caption{Density distribution of the swearing \protect\subref{fig:ecdf_swearing} (exponential y-scale) and negative content \protect\subref{fig:ecdf_negative} ratios for each user and dataset.}\label{fig:ecdf_e1}
\end{figure}

As shown in Figure~\ref{fig:ecdf_swearing}, most of the users in D2 have a low swearing ratio, that is, the number of times a tweet contains a keyword divided by the user's total number of tweets. This may be due to the fact that most of the Twitter accounts reported by Anonymous did not use English as the written language.

On the other hand, users from D1 and D3 exhibit a similar behaviour regarding the usage of swear words, with a rather similar number of users having a high swearing ratio. This suggests the inability of swear words to characterize individuals in risk of radicalisation. However, median swearing ratios in datasets D1 and D3 were $0.039126$ and $0.001010$ respectively, so there is a significant evidence that users in risk of radicalisation tend to have a higher swearing ratio than random users (\emph{Wilcoxon rank sum test}: $W=8261.5, n_1=112, n_3=120, \mbox{p-value} = 0.001025$, one-tailed).

Regarding to the use of words with negative content, it can be observed a similar effect as with swearing (see Figure~\ref{fig:ecdf_negative}): it shows a low ratio on accounts of D2, and a slightly higher density ratio on users from D1 with respect to those in D3. Again, the median negative content ratios of $0.027507$ and $0.0$ in D1 and D3, respectively, shows a statistically significant evidence that former users tend to use more words with negative connotations than the latter (\emph{Wilcoxon rank sum test}: $W=9142.5, n_1=112, n_3=120, \mbox{p-value} = \expnumber{3.715}{-07}$, one-tailed).

These results suggest that analysing the use of words with negative connotations as well as swearing, are good metrics to measure the frustration of an individual regarding his/her risk of being radicalised. In other words, it seems that anyone in risk of radicalisation is more likely to use negative and swearing words than the average twitter user. However, this should not be taken as a binary decision, but it could be used jointly with other risk radicalisation components to generate a more accurate prediction.

\newpage
\subsection{Introversion}
To measure the introversion of the user we defined two metrics: the ratio of tweets that contain an \textit{ellipsis} and the \textit{median length} of the user's tweets, computed as the number of characters and hence limited to 140 characters.

\begin{figure}[ht]
	\centering
	\subfloat[][]{\includegraphics[width=.5\textwidth]{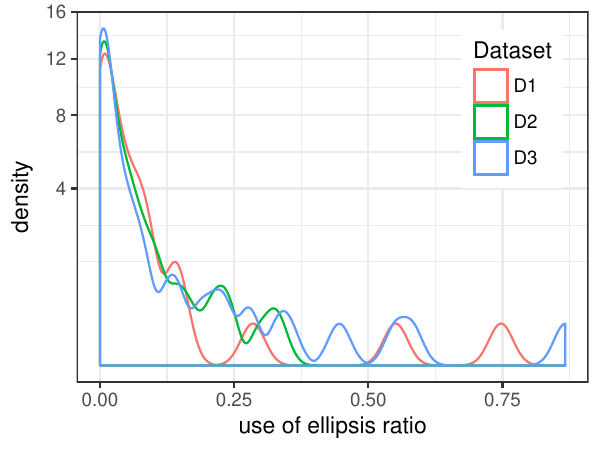}\label{fig:ellipsis}}
	\subfloat[][]{\includegraphics[width=.5\textwidth]{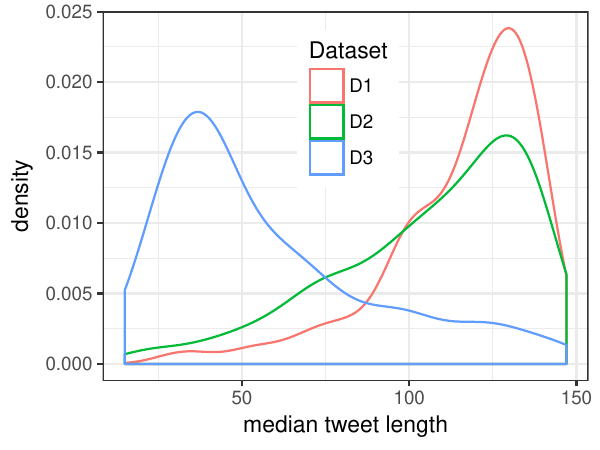}\label{fig:twitter_length}}
	\caption{Density distribution of the ellipses ratio \protect\subref{fig:ellipsis} and median tweet length \protect\subref{fig:twitter_length} for each user and dataset (exponential scale on both y-axis).}\label{fig:density_i2}
\end{figure}

Figure~\ref{fig:ellipsis} shows the distribution of the ellipsis ratio for every user and dataset. It seems that the three distributions have the same behaviour and then the ratio should not be used to measure the indicator. In fact, we were unable to find significant evidence that the ellipsis ratio was be defined by different probability distributions (\emph{Wilcoxon two-tailed rank sum test}: $W_{12}=7933, W_{13}=6777.5, W_{23}=8600.5, n_1=112, n_2=142, n_3=120, \mbox{p-values}=\left\{ 0.9744, 0.9105, 0.8952 \right\} $). This result makes sense as it is really common to use ellipsis when writing messages in a Social Network.

We found an interesting result regarding the median tweet length per user (see Figure~\ref{fig:twitter_length}): contrary to what the indicator suggests, users of the random dataset D2 write tweets shorter than those at risk of radicalisation (dataset D1 and D2), with medians $43.5, 123, 114$ respectively. Furthermore, this difference in the median length of the tweets is statistically significant according to one-tailed Wilcoxon rank sum tests with $W_{31}=1355, W_{32}=2526.5, \mbox{p-values}<0.05$.

These results suggest refocusing those metrics associated with the introversion indicator \textbf{I2}, since they are not able to establish a difference between users at risk of radicalisation and random users (use of ellipsis) or to measure the expected value (length of tweets). This makes sense: due to the limited length of tweets (140 characters), users tend to write many ellipses as a short way of enumerating things instead of writing them. Regarding the length of the tweet, as opposite to the expected, users in risk of radicalisation write longer tweets than the average user, pointing out the necessity of refocusing this metric.

\subsection{Perception of discrimination for being Muslim}
The metric associated to this indicator is similar to the metric of swearing in the case of \textbf{I1}, as both metrics counts the number of tweets using a set of precise keywords (see Table \ref{tbl:keywords}). As in the latter metric, this value is calculated using the number of times a tweet contains a particular keyword, divided by the user's total number of tweets.

\begin{figure}[ht]
	\centering
    \includegraphics[width=\textwidth]{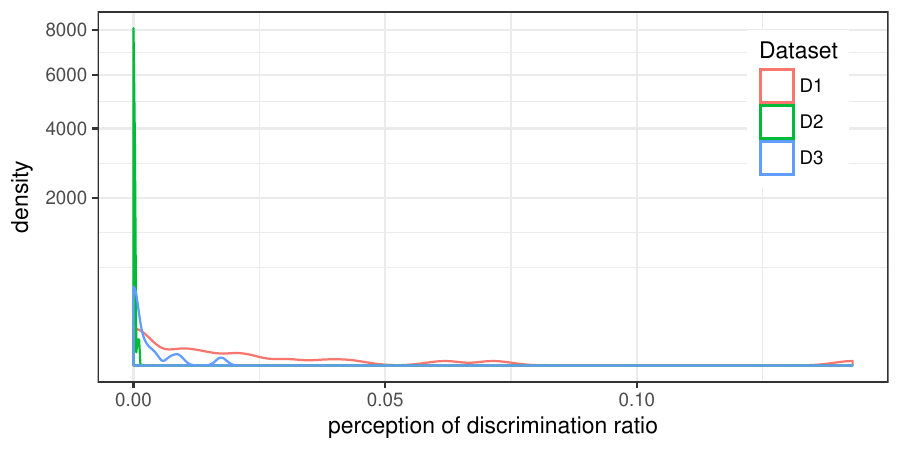}
    \caption{Density distribution of the perception of discrimination ratio (exponential y-scale).}\label{fig:discrimination}
\end{figure}

As it can be seen in Figure~\ref{fig:discrimination}, perception of discrimination for users of dataset D2 is heavily skewed to the left, that is, no tweets with discrimination keywords for almost every user. Again, a possible explanation to this observed effect is the wide range of written languages present in the dataset.

On the other hand, the ratio for dataset D1 is slightly higher than the ratio for dataset D3. This difference has been tested for statistical significance using a one-tailed Wilcoxon rank sum test with $W=8551.5, \mbox{p-value}<0.05$, even though both median are 0. This result emphasise the ability of this metric to distinguish the perception of discrimination for radicalised users.

\subsection{Negative ideas about Western Society}
\begin{figure}[ht]
	\centering
    \includegraphics[width=\textwidth]{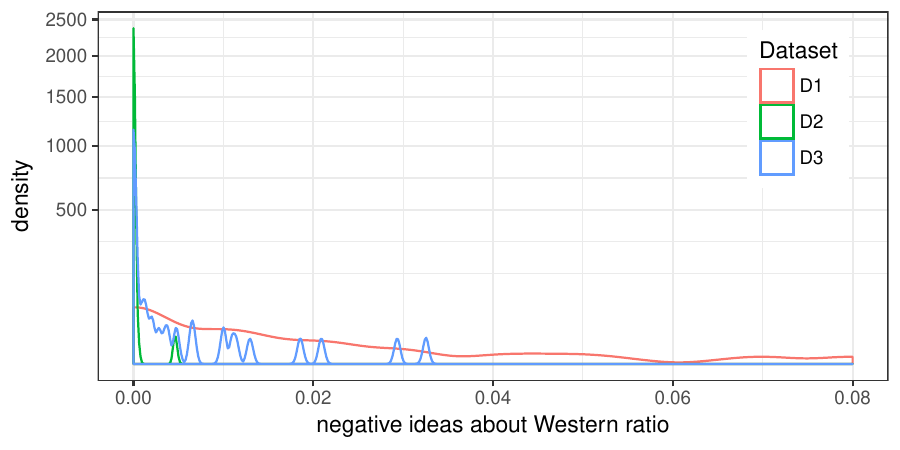}
    \caption{Density distribution of the ratio of tweets expressing negative ideas about Western society (exponential y-scale).}\label{fig:western}
\end{figure}

As occur with the previous indicator, we used a set of keywords to measure the ratio of tweets expressing negative ideas about Western society for every user and dataset (see Figure~\ref{fig:western}).

Regarding this metric we found the same results as with the discrimination indicator. The ratio of tweets with negative ideas from users in dataset D2 distributed around $0$ as expected, since the metric is keyword based. In a similar way, it is more likely to get a higher ratio if the user belong to those considered as radicalised ($W=8540, \mbox{p-value}<0.05$), thus supporting the use of this metric as a numeric feature of \textbf{I4}.

\subsection{Positive ideas about Jihadism}
\begin{figure}[ht]
	\centering
    \includegraphics[width=\textwidth]{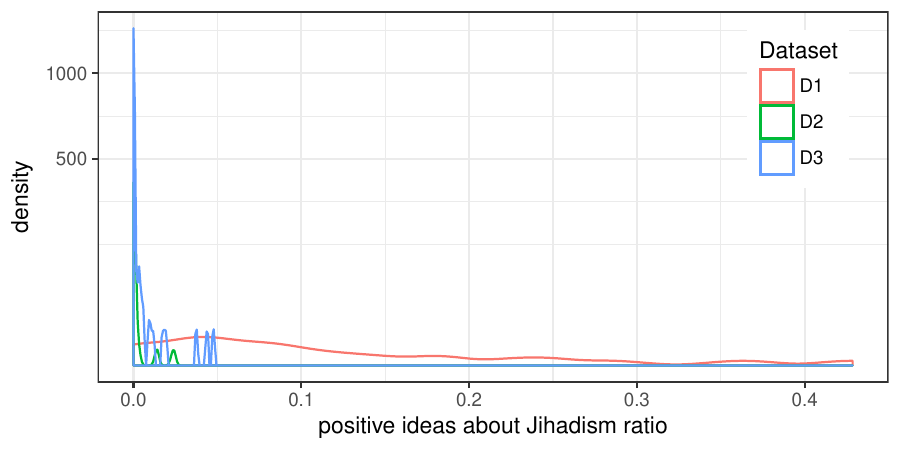}
    \caption{Density distribution of the ratio of tweets expressing positive ideas about Jihadism (exponential y-scale).}\label{fig:jihadism}
\end{figure}

Finally, analogously to the metric of indicator \textbf{I4}, we measured the ratio of tweets per user that expressed positive ideas about Jihadism. Figure~\ref{fig:jihadism} shows the results for the analysis of this metric. There is not much more to say about these results that have not been said previously: due it is a metric based on English keywords, it keeps failing at analysing tweets written in other languages (as Arabic), while there is a significant difference between the distribution of ratios for users in D1 and D3 ($W=12265, \mbox{p-value}<0.05$) which means that any user in risk of radicalisation is more willing to express positive ideas about Jihadism than a random user.

\subsection{Analysis of the ratios}
\begin{figure}[ht]
	\centering
    \includegraphics[width=\textwidth]{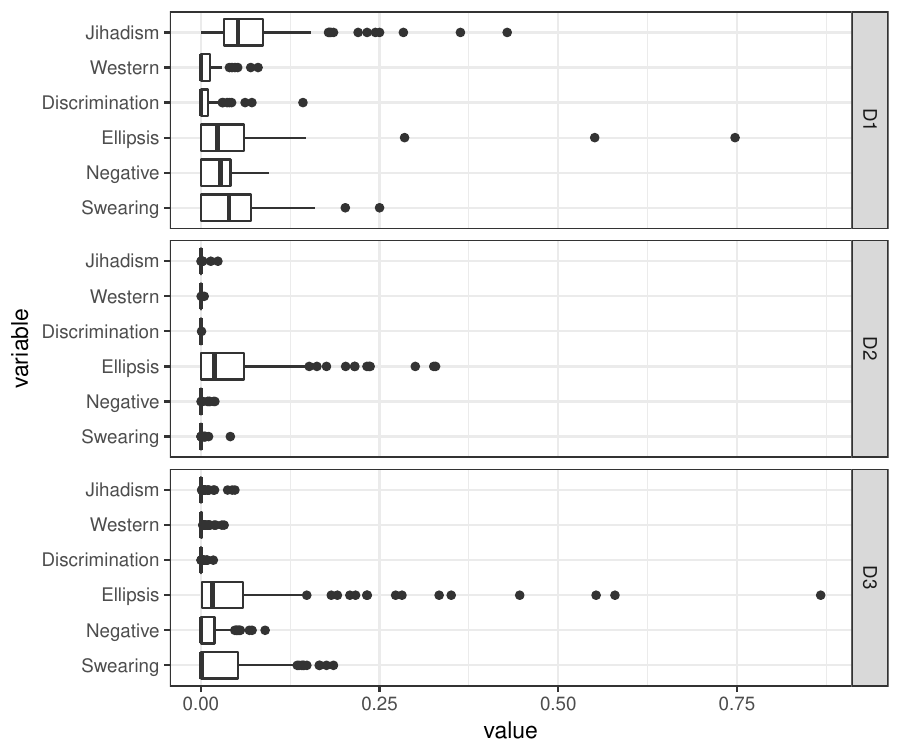}
    \caption{Distribution of the different ratios (i.e. metrics) per dataset.}\label{fig:boxplots}
\end{figure}

In order to discover the importance that each ratio has within the datasets, it have been studied their distribution over each dataset (see Figure~\ref{fig:boxplots}). As already explained, tweets from dataset D2 are written in many different languages, and those written in English represent a small part of the total. So it is not surprising that those keyword-based metrics have such low variability and their value is virtually zero. The only exception is the use of ellipsis, which seems to be a common practice in languages other than English as shown by the corresponding box-plot.

Focusing on the third dataset, which was made by the tweets from randomly selected users, there are three keyword-based metrics whose values are distributed on the low range. This makes sense as they are related to the keywords that express positive ideas about Jihadism, negative ideas about Western society and perception of discrimination, that is, they are related to very specific topics. On the other hand, ellipsis, swearing and negative content have more presence on the dataset, which suggests that is really common using and expressing them when writing messages on Social Networks.

In the case of the first dataset, the use of ellipsis follows the same behaviour as in the other datasets. As it was previously described, it is a fairly common procedure. The metric related to Jihadism keywords seems to have higher ratios than the other metrics when the user has been radicalised. This observation points out at the importance this metric should have when computing an aggregated metric to measure the risk of radicalisation.

\section{Conclusion and future works} \label{sec:conclusions}
This paper presents a study on a set of indicators devoted to risk radicalisation assessment, and their respective metrics used to quantify them. These indicators come from several experts psychologists on radicalisation, and are focused on the behaviour expressed in Social Networks. Furthermore, they can be easily computed in an automated way, which is an improvement over the current methods of risk assessment that are based on manual tools.

To study the performance of the aforementioned metrics, three datasets have been used: D1 that includes Twitter usernames and tweets from ISIS sympathizers, D2 that combines users and tweets flagged by Anonymous volunteers as radicalised people, and finally D3 that represents a random sample extracted from standard Twitter users. Precisely, we computed the metrics associated to the aforementioned indicators and then studied their density distribution. Then we compared the distributions of the datasets to discover if there is any statistically significant difference.

Regarding the ability of these metrics to highlight radicalised users, they perform generally well. We found statistical evidence that it is more likely to achieve a higher ratio of swearing, using words with negative connotations, perceiving discrimination and expressing positive and negative ideas about Jihadism and Western society, respectively, if the user is radicalised or in risk of radicalisation. Moreover, we found that radicalised users tend to write longer tweets than the rest, contrary to what was expected according to the introversion indicator.

On the other hand, we found that using ellipsis in the written text is not a relevant feature to use it as a metric for measuring introversion. This is a relevant result because this indicator has been traditionally considered as relevant in other areas as psychology \cite{merchant2001syntax,PHILLIPS201478}. Also, the performance of these keyword-based metrics were found to be extremely dependent of the language in which users wrote their tweets, so it is mandatory to increase the set of keywords with their corresponding translation in additional languages. This suggests that \textbf{I2} needs to be refocused in order to be as useful as the rest of the indicators.

Although the metrics studied in this paper have shown promising results as well as an adequate performance, they rely mainly on the messages and several set of keywords. It would be very interesting, as a possible future line of work, to define and analyse additional metrics based on the features and structure of complex networks, as Social Networks are considered of this kind.

The high dependency to the language exhibited by the indicators also lead us to think about integrating an ontology~\cite{guarino1998formal} to decouple those keywords expressed in a precise language and use language-agnostic concepts instead. Other possibility could be use directly a multilingual approach~\cite{Garcia2015,berard2016multivec} for keywords and word processing based on the most relevant languages, as Arab, English, Russian, and French. This way, indicators will be able to operate on a wider range of messages, even capturing the use of slang or additional complex relationship between the expressed ideas.

Nevertheless, once the statistical suitability of the indicators have been tested, the next step is using these indicators as features for a Machine Learning algorithm that should be able to classify a Twitter user as being in risk of radicalisation by aggregating and combining the aforementioned indicators.

\section*{Acknowledgements}
  This work has been co-funded by the following research projects: EphemeCH (TIN2014-56494-C4-4-P) Spanish Ministry of Economy and Competitivity, under the European Regional Development Fund FEDER, and Justice Programme of the European Union (2014-2020) 723180 -- RiskTrack -- JUST-2015-JCOO-AG/JUST-2015-JCOO-AG-1. The contents of this publication are the sole responsibility of their authors and can in no way be taken to reflect the views of the European Commission. Finally, we would like to thank you to the reviewers and the Editor in Chief, for the different suggestions and comments made to this work.

\section*{References}

\bibliography{main}

\end{document}